\documentclass[preprintnumbers,amssymb]{revtex4}
\usepackage{amssymb}
\usepackage{amsmath}
\usepackage{graphicx}
\usepackage{subfig}
\usepackage{dcolumn}
\usepackage{bm}

\begin{document}
\addtolength{\voffset}{-.75cm}
\addtolength{\textheight}{1.7cm}
\addtolength{\hoffset}{-0.25cm}
\addtolength{\textwidth}{.55cm}
\title{Effects of  CP violation from  Neutral Heavy Fermions on Neutrino Oscillations, and the LSND/MiniBooNE Anomalies}
\author{ Ann E. Nelson }
\affiliation{\\
 Department of Physics, Box 1560, University of Washington,
           Seattle, WA 98195
}
\newcommand{\geff}{g_{\rm eff}}
\newcommand{\meff}{m_{\rm eff}}
\newcommand{\p}[0]{\partial}
\renewcommand{\d}[0]{\textrm{ d}}
\newcommand{\R}[0]{\mathbb{R}}
\newcommand{\C}[0]{\mathbb{C}}
\newcommand{\bra}[1]{\big<#1\big|}
\newcommand{\ket}[1]{\big|#1\big>}
\renewcommand{\matrix}[4]{\left( \begin{array}{c c} #1 & #2 \\ #3 & #4 \end{array} \right)}
\newcommand{\pf}[2]{\frac{\partial #1}{\partial #2}}
\newcommand{\df}[2]{\frac{\textrm{d}#1}{\textrm{d}#2}}
\newcommand{\be}[0]{\begin{equation*}}
\newcommand{\ee}[0]{\end{equation*}}
\newcommand{\sfrac}[2]{\textstyle{\frac{#1}{#2}}}
\newcommand{\Res}[0]{\textrm{Res}}
\newcommand{\lc}[1]{\epsilon_{#1}}
\newcommand{\del}[0]{\nabla}
\newcommand{\braket}[2]{\big<#1\big|#2\big>}
\newcommand{\beq}{\begin{eqnarray}}
\newcommand{\eeq}{\end{eqnarray}}
\newcommand{\nn}{\nonumber}
\newcommand{\bml}{ $U(1)_{\textrm{\tiny{B-L}}}$}
\def\ltap{\ \raise.3ex\hbox{$<$\kern-.75em\lower1ex\hbox{$\sim$}}\ }
\def\gtap{\ \raise.3ex\hbox{$>$\kern-.75em\lower1ex\hbox{$\sim$}}\ }
\def\CO{{\cal O}}
\def\CL{{\cal L}}
\def\CM{{\cal M}}
\def\tr{{\rm\ Tr}}
\def\CO{{\cal O}}
\def\CL{{\cal L}}
\def\CM{{\cal M}}
\def\mpl{M_{\rm Pl}}
\newcommand{\Dslash}{D\llap{/\kern+2.1pt}}
\newcommand{\bel}[1]{\be\label{#1}}
\def\al{\alpha}
\def\bt{\beta}
\def\eps{\epsilon}
\def\eg{{\it e.g.}}
\def\ie{{\it i.e.}}
\def\mn{{\mu\nu}}
\newcommand{\rep}[1]{{\bf #1}}
\def\be{\begin{equation}}
\def\ee{\end{equation}}
\def\bea{\begin{eqnarray}}
\def\eea{\end{eqnarray}}
\newcommand{\eref}[1]{(\ref{#1})}
\newcommand{\Eref}[1]{Eq.~(\ref{#1})}
\newcommand{\gsim}{ \mathop{}_{\textstyle \sim}^{\textstyle >} }
\newcommand{\lsim}{ \mathop{}_{\textstyle \sim}^{\textstyle <} }
\newcommand{\vev}[1]{ \left\langle {#1} \right\rangle }
\newcommand{\ev}{{\rm eV}}
\newcommand{\kev}{{\rm keV}}
\newcommand{\Mev}{{\rm MeV}}
\newcommand{\gev}{{\rm GeV}}
\newcommand{\tev}{{\rm TeV}}
\newcommand{\mev}{{\rm MeV}}
\newcommand{\mnu}{\ensuremath{m_\nu}}
\newcommand{\mlr}{\ensuremath{m_{lr}}}
\newcommand{\acc}{\ensuremath{{\cal A}}}
\newcommand{\mav}{MaVaNs}
\newcommand{\nusm}{$\nu$SM }

\begin{abstract}
Neutrinos  may mix  with ultralight fermions, which   gives flavor oscillations, and  with heavier fermions, which yields short distance flavor change.  I consider the case where both effects are present.
  I show that in the limit where a single oscillation length is experimentally accessible, the effects of  heavier fermions on neutrino oscillations can generically be accounted for by a simple  formula containing four parameters,  including observable  CP violation. I consider the anomalous LSND and MiniBooNE results, and show that these can be fit in  a model with CP violation and two additional sterile neutrinos, one in the   mass range between  0.1 and 20 eV, and  the other with mass between 33 eV and 40 GeV.   I also show that this model can avoid conflict with   constraints from existing null short baseline experimental results.  

\end{abstract}
\maketitle
\section{Introduction}
Since the discovery of neutrino flavor change in a variety of long baseline experiments\cite{Casper:1991ac,Becker-Szendy:1992hq,Hirata:1992ku,Fukuda:1994mc,Fukuda:1996sz,Hampel:1998xg,Fukuda:1998mi,Cleveland:1998nv,Abdurashitov:1999zd,Eguchi:2002dm,Ahmad:2001an,Ahmad:2002jz,Ahmed:2003kj}, a new standard picture has emerged \cite{Donini:1999jc,Maltoni:2002ni,review,deGouvea:2004gd,Mohapatra:2005wg,Fogli:2005cq,Strumia:2006db,Mohapatra:2006gs,GonzalezGarcia:2007ib,Schwetz:2008er,Fogli:2009zza}.
In this picture the   neutrino flavor eigenstates $e, \mu, \tau$ are related to the mass eigenstates $\nu_{1,2,3}$ via a 3-by-3 unitary matrix \cite{Maki:1962mu}:
\be\label{stdosc}
\left( \begin{array}{c}
\nu_e \\
\nu_{\mu} \\
\nu_{\tau}
\end{array} \right) = \left( \begin{array}{c c c}
1 & 0 & 0 \\
0 & c_{23} & s_{23} \\
0 & -s_{23} & c_{23}
\end{array} \right)\left( \begin{array}{c c c}
c_{13} & 0 & s_{13}e^{-i\delta} \\
0 & 1 & 0 \\
-s_{13}e^{i\delta} & 0 & c_{13}
\end{array} \right)\left( \begin{array}{c c c}
c_{12} & s_{12} & 0 \\
-s_{12} & c_{12} & 0 \\
0 & 0 & 1
\end{array} \right)\left( \begin{array}{c}
\nu_1 \\
\nu_2 \\
\nu_3
\end{array} \right)\ ,
\ee
where $c_{ij} = \cos\theta_{ij}$, $s_{ij} = \sin\theta_{ij}$.   Long  baseline measurements are consistent with the following values of the three angles $\theta_{ij}$:
\be\label{stdangles}
\begin{split}
\tan^2 \theta_{12} &= 0.45\pm0.05 \\
\sin^2 2\theta_{13} &= 0^{+0.05}_{-0} \\
\sin^2 2\theta_{23} &= 1.0^{+0}_{-0.1}
\end{split}
\ee
and  neutrino mass squared differences:
\be\label{stdmasses}
\begin{split}
\Delta m_{12}^2 &= (8.0\pm0.3)\cdot10^{-5}\textrm{ eV}^2 \\
\big|\Delta m_{23}^2\big| &= (2.5\pm0.2)\cdot10^{-3}\textrm{ eV}^2
\end{split}
\ee

These mass eigenstates are so light and nearly degenerate that in all neutrino experiments neutrinos  propagate at essentially the speed of light, and the  components of the neutrino wavepackets with different mass do not separate spatially.  The phases of the different mass components oscillate quantum mechanically with different frequencies, and, as the different flavors are different superpositions of mass eigenstates, the flavor composition of a neutrino beam in vacuum will exhibit spatial variation  as a function of $L/E$, where $E$ is the neutrino energy and $L$ is the distance from the source \cite{Pontecorvo:1967fh}.  In propagation through matter the phases are also altered by forward scattering  from  the weak interactions, which alters the  
flavor change probability in a way which depends on both $E$ and $L/E$, and differs for neutrinos and anti neutrinos \cite{Wolfenstein:1977ue,Mikheev:1986gs}. These matter effects are very small in experiments with baseline much shorter than 1000 km,  unless there exist exotic  forces \cite{Valle:1987gv,Roulet:1991sm,Bergmann:1999rz,Chang:2000xy,Grifols:2003gy,GonzalezGarcia:2006vp,Nelson:2007yq}.  Oscillations are now significantly favored over  alternatives such as neutrino decoherence or decay \cite{Ashie:2004mr,Adamson:2008zt}.

The  measurement of the small mixing angle $\theta_{13}$ is a primary goal of the current generation of long baseline experiments. Also sought is evidence for a nonzero value of the  CP  violating parameter $\delta$, and knowledge of whether the pair of states with  the smaller mass squared splitting $\Delta m_{12}^2 $ are heavier or lighter than the third state.

There have  been reports of neutrino flavor change in the short baseline LSND \cite{Aguilar:2001ty}  and MiniBooNE \cite{AguilarArevalo:2007it,AguilarArevalo:2008rc,AguilarArevalo:2010wv}  experiments, which would upset this standard picture, as the values of $L/E$ in both these experiments is of order 1 MeV/m, which is too small for the small mass squared splittings $\Delta m_{12}^2$  and $\Delta m_{23}^2$ to effect     flavor change at the observed level.  The LSND and MiniBooNE results  favor anti electron neutrino appearance in a muon antineutrino beam, at different energies and distance but similar values of $L/E$. The   MiniBooNE data on neutrinos disfavors electron neutrino appearance in a muon neutrino beam at the values of $L/E$ explored by LSND, but favors an excess of electron neutrinos at higher values of $L/E$ \cite{AguilarArevalo:2008rc}. With at least 2 additional sterile neutrinos, CP  violation in oscillations can reconcile the LSND and MiniBooNE anti neutrino results  with MiniBooNE neutrino results \cite{Maltoni:2007zf,Karagiorgi:2009nb,Karagiorgi:2010zz}. However reconciling   MiniBooNE and LSND with constraints on muon or electron neutrino disappearance from a variety of other short baseline experiments \cite{Dydak:1983zq,Stockdale:1984ce,Stockdale:1984cg,Declais:1994su,Apollonio:1999ae} is more difficult. Attempts to do so have introduced additional exotic ingredients beyond neutrino mixing 
\cite{Pas:2006tk,Maltoni:2007zf,Nelson:2007yq,Schwetz:2007cd,Farzan:2008zv,Arias:2009fk,Diaz:2009qk,Hollenberg:2009tr,Gninenko:2009ks,Giunti:2009zz,Akhmedov:2010vy}.

Existing studies of neutrino oscillations generally are not sensitive to mass squared differences larger than 1000~eV$^2$, as the resulting oscillation length is too short to measure. Furthermore mixing with such heavy neutrinos has not previously been considered as a mechanism to reconcile LSND and MiniBooNE with short baseline disappearance constraints.
In this paper we will consider electron neutrino or anti neutrino appearance in a muon neutrino or anti neutrino beam in the case where at least one   neutrino is so heavy that the associated mass squared difference gives rise to an unobservably short oscillation length.

 \section{Neutrino Oscillations and  mixing  with a    neutrino heavier than  33 eV}
In this section we will assume  neutrino mass differences which are are small enough so that the difference mass components of the wavefunction do not separate spatially. In the appendix I review the derivation of neutrino oscillation formulae using a wavepacket formalism, and give bounds on the length scale where this assumption to be valid. For instance for the MiniBooNE experiment the mass difference must be less than about  20 keV. Assuming a unitary mixing matrix,  the  formula for the probability of electron  neutrino appearance in a muon neutrino beam is then given by the standard result
\be  = \left|\sum_{i>1}U_{ei}U_{\mu i}^*( e^{-2 i x_{i1}} -1)\right|^2  \ee
where \be x_{ij}\equiv 1.27\frac{ (m_i^2-m_j^2)}{\rm eV^2}\frac{L/E}{\rm  m/MeV} \ . \ee 
Specializing to the short baseline case, where the mass differences among the three light states can be neglected, and assuming two additional states, this formula becomes
\be P_{\nu_\mu\rightarrow \nu_e } = \left|U_{e4}U_{\mu4}^* e^{-2 i x_{41}} + U_{e5}U_{\mu 5}^* e^{-2ix_{51}}-U_{e 4}U_{\mu4}^*-U_{e5} U_{\mu5} ^*\right|^2  \ .\ee
 For anti neutrinos the   matrix elements are  complex conjugated.
This probability can be written as  \cite{Kayser:2002qs}
\be P_{\nu_\mu\rightarrow \nu_e }= 4 |U_{e 4}|^2 |U_{\mu 4}|^2\sin^2 x_{41} + 4 |U_{e 5}|^2 |U_{\mu 5}|^2\sin^2 x_{51} +8|U_{e 5}||U_{e4}||U_{\mu 4}| |U_{\mu 5}|\sin x_{51}\sin x_{41}\cos(x_{51}-x_{41}-\phi)\ee
where  \be\phi\equiv \arg\left( \frac{U_{e5}U_{\mu 5}^*}{U_{e 4}U_{\mu 4}^*}\right)\ee is a physically observable CP violating phase and  for  antineutrinos we must replace $\phi-\rightarrow -\phi$.
In the limit where the 5th neutrino is   heavy $x_{51}$ varies very rapidly and should be averaged over.  The appearance probability then becomes
\bea&& P_{\nu_\mu\rightarrow \nu_e }4 |U_{e 4}|^2 |U_{\mu 4}|^2\sin^2 x_{41} + 2 |U_{e 5}|^2 |U_{\mu 5}|^2+4|U_{e 5}||U_{e4}||U_{\mu 4}| |U_{\mu 5}|\sin x_{41}\sin(x_{41}+\phi)  \ .\eea

We can simplify this expression by  defining the CP odd quantity $\beta$
\be\beta\equiv \frac{1}{2} \tan^{-1}\left(\frac{\sin\phi |U_{e 5}||U_{\mu 5}| }{  |U_{e 4}| |U_{\mu 4}|+\cos\phi |U_{e 5}||U_{\mu 5}| }\right)\ee
and the mixing ratio $r$
\be r\equiv \frac{|U_{e 5} U_{\mu 5}^* +  U_{e 4} U_{\mu 4}^*|}{|U_{e 4} U_{\mu 4}^*|},\ee
so that
\be  r e^{ 2 i \beta}=\frac{U_{e5}U_{\mu5}^*+U_{e4}U_{\mu 4}^*}{U_{e4}U_{\mu 4}^*} \ee
and get oscillation probability
\bea P_{\nu_\mu\rightarrow \nu_e }&=&  |U_{e 4}|^2 |U_{\mu 4}|^2[2(1-r)^2 + 4 r \sin^2\beta+ 4r\sin^2(x_{41}+\beta) ]
 \ . \eea For anti neutrinos we replace $\beta\rightarrow-\beta$. Note that CP violation remains observable in the limit of heavy $m_5$. Note also that while with 4 neutrinos the   amplitude  of oscillations associated with $m_4$ is $4 |U_{e 4}|^2 |U_{\mu 4}|^2$,   mixing with a heavy 5th neutrino alters the oscillation amplitude by a factor of $r$, which may be either larger or smaller.

In contrast with appearance  experiments, in the absence of large matter effects, disappearance experiments are much less sensitive to mixing with heavy neutral fermions. Averaging over the short oscillation length associated with  $\Delta m_{51}^2$, and neglecting the mass differences among the three light neutrinos, the probability for vacuum electron  neutrino or electron anti neutrino disappearance is
\be4 |U_{e4}|^2(1-|U_{e4}|^2-|U_{e5}|^2)\sin^2x_{41}+ 2|U_{e5}|^2(1-|U_{e5}|^2)\ee
and the probability for muon neutrino or muon antineutrino disappearance is obtained by replacing $e\rightarrow\mu$ in the preceding formula.    Typically, stringent bounds  on disappearance are obtained by canceling systematic errors using  the $L/E$ dependence.  Thus  $U_{e5}$ and $U_{\mu5}$ are only   weakly constrained   in the limit where  the oscillation length associated with $\Delta m_{51}^2$ is too short to give any measurable  dependence on $L/E$.  

The tension between electron antineutrino appearance at  LSND and MiniBooNE  and short baseline electron and muon neutrino disappearance experiments may be reduced by allowing both nonzero $\beta$ and $r$    greater than 1.  The ratio $r$ is constrained by very short baseline electron neutrino appearance searches. The strongest such  constraints come  from the NOMAD and E776 experiments \cite{Borodovsky:1992pn,Astier:2003gs} .  The NOMAD constraint on $\nu_e$ appearance at $L/E <0.025$m/MeV  implies
\be  |U_{e 4}|^2 |U_{\mu 4}|^2((1-r)^2 + 4 r \sin^2\beta )<0.0007\ .\ee  
For instance for $\beta=0$, and $|U_{e 4}|^2 |U_{\mu 4}|^2=2\times 10^{-4}$, the maximum  allowed value of $r$ is 3. For a 3+1 model with $\delta=0 $ and $r=1$, the maximum probability of electron neutrino appearance as a function of $L/E$ is $ 8 \times10^{-4}$.  In contrast, for  $r=3$, the probability of electron neutrino appearance maximizes  at   a much larger $4\times 10^{-3}$. Therefore   the existence of the heavy fifth neutrino makes it is possible to obtain much larger electron neutrino or antineutrino appearance probabilities at non zero $L/E$ than would   be possible in a 3+1 model.  In the next section we will find even larger  values of $r$ are possible when neutrinos mix  with fermions which are too heavy to be produced.

 \section{A heavy neutrino}
   In this section we consider in detail a model with 2 additional sterile neutrinos, one of which  is heavy enough  that it is possible to kinematically distinguish it from the others. In principle it  is produced with a reduced phase space, or, if sufficiently heavy, it is not produced at all. Mixing with a heavy neutrino will be constrained from loop contributions to charged lepton flavor violation, such as $\mu\rightarrow e \gamma$ \cite{Antusch:2006vwa}, but as long as it is lighter than about 40 GeV      such a neutrino could contribute substantially to the LSND/MiniBooNE anomalies while the charged lepton flavor violation rate will be  sufficiently suppressed by the Glashow Iliopoulos Maiani  \cite{Glashow:1970gm} mechanism.
In the case of neutrino mixing with such a heavy neutrino, the  beam will  not initially be in a pure flavor eigenstate, and the situation can be described in terms of a non-unitary mixing matrix for the light states\cite{Antusch:2006vwa}. Without the unitarity constraint, CP violation is possible even in two neutrino oscillations \cite{FernandezMartinez:2007ms}.  In this section we will consider the effect of an additional state which is significantly heavier than the other neutrinos, but light enough so that charged lepton constraints on unitarity violation are not constraining.
 
 Neglecting the mass differences among the 3 light eigenstates,  but assuming there is a 5th heavy neutrino and a 4th neutrino whose mass squared difference with the others is not negligible, the formula for the probability of electron neutrino appearance in a muon  neutrino beam is  

\bea
P_{\nu_\mu\rightarrow \nu_e  } &=&\left|U_{e4}U_{\mu4}^* e^{-2 i x_{14}} -U_{e 4}U_{\mu4}^*-U_{e5} U_{\mu5}^* \right|^2 +  a   |U_{e 5}|^2 |U_{\mu5}|^2\\
&=&  |U_{e 4}|^2 |U_{\mu 4}|^2|e^{-2 i x_{14}} -r e^{  2 i \beta}|^2 +   a   |U_{e 5}|^2 |U_{\mu5}|^2\ ,
\eea
giving  electron neutrino appearance  probability
\be P_{\nu_\mu\rightarrow \nu_e  }  = |U_{e 4}|^2 |U_{\mu 4}|^2\left\{(1-r)^2+ a[(1-r)^2+4 r \sin^2\beta] +4 r \sin^2 (x_{41}+\beta)\right\}\ , \ee with $\beta\rightarrow -\beta$ for antineutrinos. Here $a$ is a phase space factor associated with production of the heavy state, which is less than 1, and which is 0 if the state is heavier than the available energy.  In this case of $a<1$  the constant term is smaller than    in the previous section,  because the contribution from the 5th neutrino  to the constant term is reduced. Assuming that $a=0$, the constraint on $r$ from very short baseline electron neutrino appearance is correspondingly weakened, to
\be  |U_{e 4}|^2 |U_{\mu 4}|^2\left[\frac{(1-r)^2}{2} + 2 r \sin^2\beta \right]<0.0007\ .\ee For instance for $\beta=0$, and $|U_{e 4}|^2 |U_{\mu 4}|^2=2\times 10^{-4}$, the maximum  allowed value of $r$ is 3.8, and  the probability of electron neutrino appearance maximizes  at     $8\times 10^{-3}$. In contrast, in a 3+1 model with $|U_{e 4}|^2 |U_{\mu 4}|^2=2\times 10^{-4}$, and no 5th heavy neutrino, the maximum short baseline electron appearance probability would be only $8\times 10^{-4}$.  

Also somewhat constraining will be the muon decay rate, which could be affected from  $\nu_\mu$ and $\nu_e$ mixing with a sufficiently heavy neutrino, and which is constrained from lepton universality and from precision electroweak tests\cite{Marciano:1999ih}. For a   5th neutrino which is heavier than the muon, we must require $|U_{e 5}|$ and $|U_{\mu 5}|$ to be smaller than $\sim0.05$.

\section{General formula for analyzing neutrino appearance oscillation experiments in vacuum}
The results of the previous two sections are easily generalized. 
Any single neutrino oscillation experiment, is typically sensitive to oscillations in a range of $L/E$ which varies by no more than an other of magnitude or so. This suggests a simple generalization of the 2 flavor dominance formula.  Assuming a single oscillation length is comparable to the range of the experiment,  much shorter oscillation lengths may be averaged over, and much longer oscillation lengths may be neglected, the probability $P_{a\rightarrow b}$ of appearance of flavor $a$ in a beam of flavor $b$  in vacuum may generically be written
\be \label{master} P_{a\rightarrow b}=\sin^2(2\theta_{ab}) \sin^2\left( 1.27 \frac{\Delta m^2 L/E}{\rm eV^2\ m/MeV} +\beta\right)+.5\cos^2(2\theta_{ab})\sin^2\alpha\ ,\ee where $\Delta m^2$ is the relevant mass squared difference, $\theta_{ab}$ is an effective mixing angle, $\beta$ is a CP violating  phase difference between the two different components (which is allowed to be nonzero if we do not have two flavor unitarity), and $\alpha$ gives the constant term resulting from averaging over  short wavelength oscillations, from any nonunitarity,  and  from any neutrinos which are too heavy to participate in oscillations. For antineutrinos the oscillation probability in this limit would be
\be P_{\bar{a}\rightarrow \bar{b}}=\sin^2(2\theta_{ab}) \sin^2\left(1.27 \frac{\Delta m^2 L/E}{\rm eV^2\ m/MeV}-\beta\right)+.5\cos^2(2\theta_{ab})\sin^2\alpha\ .\ee
This parameterization is chosen to satisfy the generic constraints for CPT conserving vacuum oscillations $0\le P_{a\rightarrow b}\le 1$ and $0\le \langle P_{a\rightarrow b}\rangle \le1/2$.
For example, in the five neutrino model of the previous section, we would have 
\bea\sin^2(2\theta_{e\mu})&=&4 |U_{e 5} U_{\mu 5}^* +  U_{e 4} U_{\mu 4}^*| |U_{e 4}U_{\mu 4}| \\
.5\cos^2(2\theta_{e\mu})\sin^2\alpha&=&  |U_{e 4}|^2 |U_{\mu 4}|^2\left\{(1-r)^2+ a[(1-r)^2+4 r \sin^2\beta]\right\}\ . \eea The CP odd parameter $\beta$ would have the same definition as in that model. Note that  $\alpha$ may be as small as $0$ in the case where $r=1$ and $a=0$, for arbitrary value of $\beta$.
\section{LSND and MiniBooNE}

The Liquid Scintillator Neutrino Detector (LSND) experiment at Los Alamos \cite{Aguilar:2001ty} has reported statistically significant  (3.8 $\sigma$) evidence for electron anti neutrinos in a beam produced by the decay of $\mu^+$ at rest, consistent with oscillations of anti muon neutrinos. The MiniBooNE experiment at Fermilab  \cite{AguilarArevalo:2007it,AguilarArevalo:2008rc,AguilarArevalo:2010wv} which has similar range of $L/E$ to  LSND, has   searched for  muon to electron neutrino and  anti neutrino appearance. The MiniBooNE electron neutrino appearance results showed no excess in the preferred analysis region but do show an excess at lower energies. The MiniBooNE anti neutrino data shows an excess which is consistent with a neutrino oscillation interpretation of  the LSND signal, and which  is poorly fit by  background. The KARMEN experiment \cite{Armbruster:2002mp} also searched for anti electron neutrino appearance in an anti muon neutrino beam, at  values of $L/E$ ranging form  0.36 to 0.74, and saw no excess, giving a 90\% Cl limit on the oscillation probability in this region of 0.0017.  Several experiments have searched for muon to electron neutrino conversion at very short baseline with results consistent with 0 \cite{Borodovsky:1992pn,Astier:2003gs}, with the  NOMAD experiment providing the strongest constraint. 

In order to test whether \Eref{master} could account for the LSND and MiniBooNE results, I have taken the oscillation probabilities given in ref. \cite{AguilarArevalo:2010wv}  for electron neutrino and antineutrino appearance as a function of $L/E$ to construct a  $\chi^2$ function
\be\chi^2(\theta_{\mu e},\Delta m^2,\alpha,\beta)\,=\,\sum_{i}\frac{(P_i^{\rm theory}(\theta_{\mu e},\Delta m^2,\alpha,\beta)\,-\,P_i^{\rm exp})^2}{ \sigma^2_{i}} \,, \ee
where $P_i^{\rm exp}$ represents the oscillation probability for  bin $i$ extracted from experimental results, $P_i^{\rm theory}(\theta_{\mu e},\Delta m^2,\alpha,\beta)$ is given by \Eref{master}, averaged over the range of $L/E$ included in bin $i$,  and $\sigma_{i}$ is the experimental error. I include 8 bins for LSND, and 9 bins each for MiniBooNE neutrinos, and for MiniBooNE anti neutrinos. I do not use the     MiniBooNE data for $E<400$ MeV ($L/E > 1.37 $  m/MeV) because of the large systematic error, which should be correlated, as I do not have access to the correlation data. Inclusion of these points with the systematic error included and treated as uncorrelated makes little difference in the fits, but is not justifiable. I also include in the fit a bin for KARMEN, and a bin for NOMAD, with the experimental errors chosen to correspond to the 90\% upper bound on the average oscillation probability. The total number of fit points included is 28, and there are 4 free parameters. 

\begin{figure}[htbp]
{\label{chisq}\includegraphics[width=.9\textwidth]{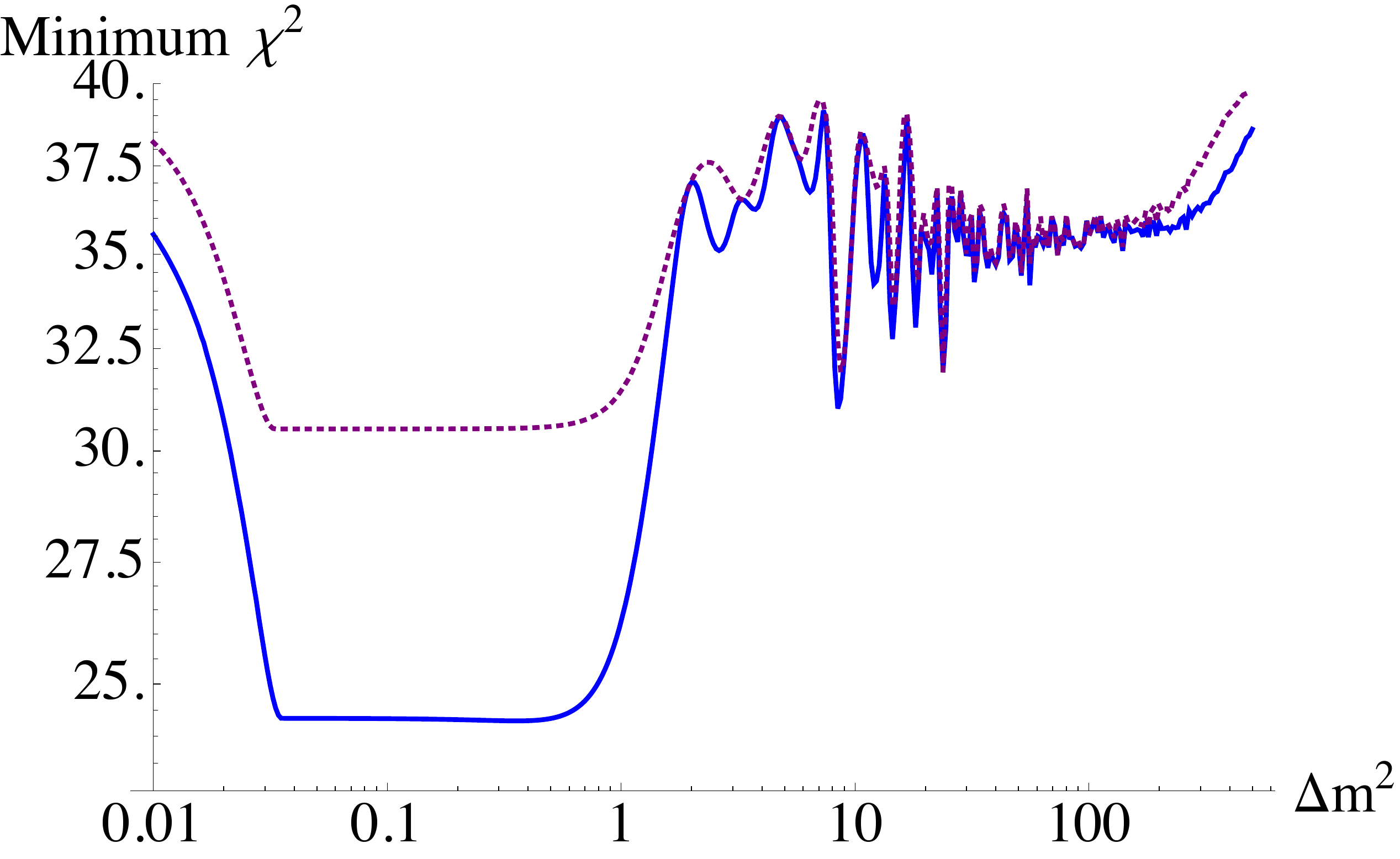}}
\caption{The  minimum  value of the $\chi^2$ function described in the text    as a function of the  $\Delta m^2_{41}$ mass squared difference in eV$^2$. The solid blue line shows the minimum with    $\theta_{\mu e}$, $\alpha$ and $\beta$ chosen to minimize $\chi^2$. The dotted purple line shows the minimum with  $\theta_{\mu e}$,  and   $\alpha$ chosen to minimize $\chi^2$ and the CP violating parameter $\beta$ set to 0, showing that the  best fit region has   a mass squared difference between  0.03 and 0.60~eV$^2$ and nonvanishing CP violation.}
\end{figure}

The best fit point has $\Delta m^2=0.40$eV$^2$, $\sin^2(2\theta_{\mu e})=0.0083$, $\beta=-0.123$,   $\alpha=0$, and a total $\chi^2$ of 24.14 for 24 degrees of freedom. A nearly equally good fit may be obtained for any $ \Delta m^2 $ in the range from 0.03 eV$^2$ to 0.60 eV$^2$. The fit  prefers a nonzero value for the CP violating parameter $\beta$. In Fig.~\ref{chisq} I show the lowest $\chi^2$ obtainable for a given value of $\Delta m^2$, with and without the $\beta=0$ constraint.  The value of $\chi^2$ at $\theta_{\mu e}=\alpha=0$ (no flavor change) is 45.6. Note that     \Eref{master} appears to give a good fit to all the data on electron neutrino or anti neutrino appearance at MiniBooNE and LSND, while being compatible with  KARMEN and NOMAD, for a wide range of masses.

\begin{figure}[htbp]
{\label{modelprobabilities}\includegraphics[width=.9\textwidth]{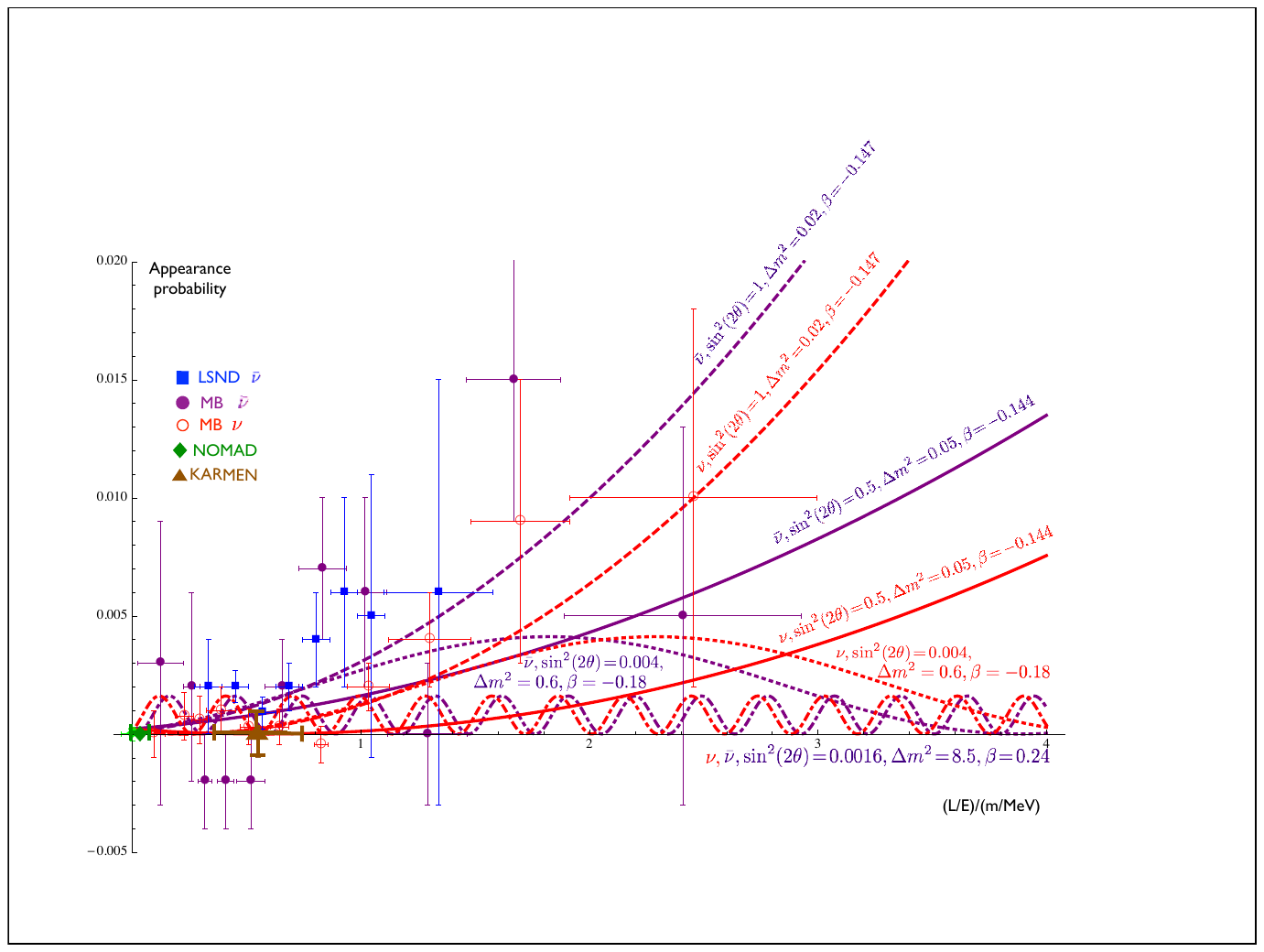}}
\caption{Electron flavor appearance probability against  $L/E$ in a muon neutrino or antineutrino beam for the  4 different parameter values indicated, all of which have $\chi^2<31$ for (24 degrees of freedom). In all cases shown the parameter $\alpha$ is zero.  The neutrino appearance probabilities are shown in red (med gray) and the anti neutrino probabilities in purple (dark gray).  The neutrino and antineutrino probabilities differ   for the same parameters due to CP violation. Also shown are the  probabilities extracted from the MiniBooNE neutrino and antineutrino data,  LSND, KARMEN   and NOMAD.}
\end{figure}

\begin{figure}[htbp]
{\label{contourplot}\includegraphics[width=.9\textwidth]{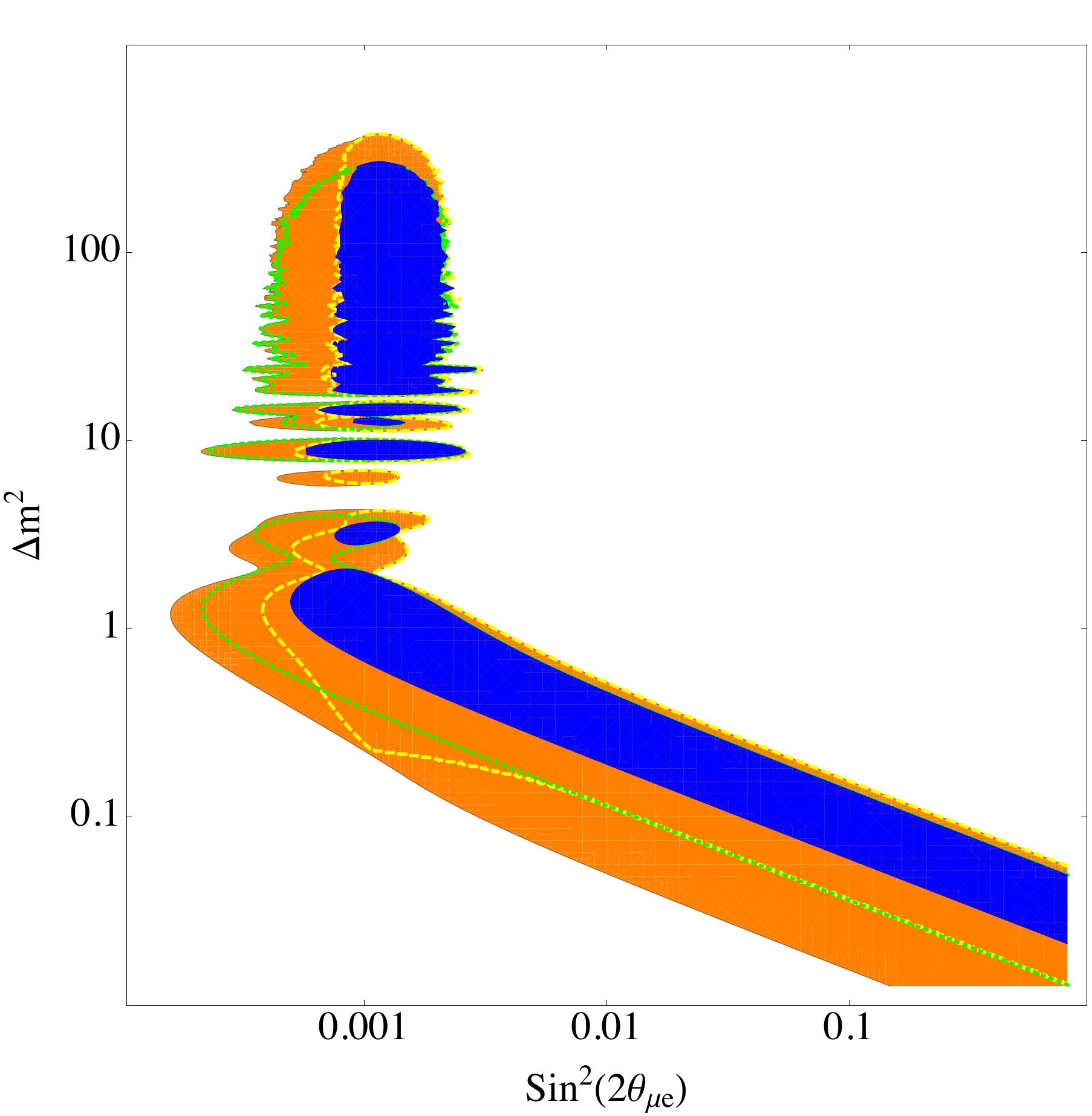}}
\caption{The   $\chi^2\le37.7$   region for   the $\Delta m^2_{41}$ mass squared difference and effective mixing angle is shown in orange (medium gray), with $\alpha$ and $\beta$ chosen to minimize $\chi^2$. Also shown  in blue (dark gray) is the preferred region  with the   parameters $\alpha$ and $\beta$ set to 0, which corresponds to a 3+1 neutrino model. The  green (light gray) dotted line shows the preferred region when $\beta=0$ (no CP violation) and $\alpha$  chosen to minimize $\chi^2$, while the yellow (very light gray) dashed line shows the preferred region when $\alpha=0$ and $\beta$   chosen to minimize $\chi^2$.}
\label{confidence}
\end{figure}

In Fig. \ref{modelprobabilities} I show the LSND and MiniBooNE electron antineutrino appearance probabilities and the MiniBooNE electron neutrino appearance probabilities as a function of L/E, together with the curves from 4 points with $\chi^2<31$. Also shown are the constraints from NOMAD and KARMEN. 
 In Fig.~\ref{confidence}, I show the region in the  $\Delta m^2$ and $\theta_{\mu e}$ plane where the  $\chi^2$ goodness of fit test is within a factor of 10 of the best  value ( $\chi^2$  less than 37.7) for four different assumptions about the $\alpha$ and $\beta$ parameters. Because the data has been extracted from the published plots without including information  about correlations these results should be taken as indicative of the preferred values rather than as a definitive constraint region.

Note that the inclusion of the $\alpha$ and $\beta$ parameters has little effect on the best fit values of $\Delta m^2$, but  greatly increases the preferred region for $\theta_{\mu e}$, allowing the effective mixing angle to be much smaller for a given $\Delta m^2$ than in a 3+1  model. I do not show the constraints on the allowed region from disappearance only experiments, since, as discussed in the previous sections, these depend on   $U_{e 5}$ and $ U_{\mu 5} $, which are only weakly constrained for some values of $m_5$. I conclude that the inclusion of CP violation and  nonunitarity in the 3+1 dimensional mixing matrix allows for mixing of the 3 active neutrinos with a  sub eV mass sterile neutrino  to fit all the short baseline electron flavor appearance data without necessarily conflicting with muon and electron neutrino disappearance data. Previous attempts  \cite{Maltoni:2007zf,Karagiorgi:2009nb,Karagiorgi:2010zz} to fit the short baseline data with oscillations among 3 active plus 2 sterile neutrinos were not able to obtain as good a fit because  only   mass squared differences of less than 1000 eV$^2$ were considered, for which the values of $U_{e 5}$ and  $U_{\mu 5} $ are more constrained.

 \section{Summary}
 Neutrino oscillation experiments   offer an unparalleled  window into exotic physics beyond the Standard Model.   
A simple extension of the standard model is to add `sterile' fermions which are neutral under all gauge interactions. The theoretical motivations for such fermions include grand unified theories,  Dirac neutrino masses, the seesaw model of neutrino mass, supersymmetric models,  dark matter theories, and exotic hidden sectors. Such fermions could mix with neutrinos, and the mixing angles are not necessarily correlated with neutrino mass.  If these exotic fermions are light they can appear in neutrino oscillation experiments as a new state, providing a an additional oscillation length. Even if they are not light, they can affect neutrino oscillations by allowing the mixing matrix among the light states to be nonunitary.  

 In this paper I considered   the existence of  neutral fermions of a wide range of masses,   which   mix significantly with neutrinos, and  showed that   in the limit of sensitivity to a single oscillation length the usual 2 parameter oscillation formula should be generalized to a 4 parameter formula, which accounts for CP violation and for a distance and energy independent component to flavor change. This formula has the advantage of covering a wide range of possible physics which can affect neutrino flavor change with relatively few parameters.
 
 I  also performed a fit of  the electron neutrino and antineutrino short baseline appearance data to the new formula, and found a good fit to the anomalous LSND and MiniBooNE results. The best fit parameter region has an additional sterile neutrino with a mass squared difference between 0.03 and 0.60 eV$^2$, and a nonunitary, CP violating mixing matrix, which could result from from mixing with a a second state, with mass between 33 eV and 40 GeV.   CP violation allows for a good fit to both the MiniBooNE neutrino results and the MiniBooNE and LSND antineutrino  results. Constructive interference with   short distance flavor change term can enhance the amplitude of the oscillatory term in appearance experiments, but not in disappearance experiments, allowing for reconciliation of the evidence for neutrino flavor change at LSND and MiniBooNE  with the lack of evidence from short baseline disappearance searches. While this fit is not conclusive evidence for new states or CP violation it is intriguing.

 \section{Appendix:Standard picture of neutrino oscillations in vacuum: wave packet formalism}
Derivations of the standard neutrino oscillation formula  have been presented many times.  In this section  I will use a   general wavepacket formalism in order to illustrate the approximations necessary for the standard  treatment, as these approximations   break down when mixing with heavier neutral fermions is considered. For simplicity I will only consider one spatial dimension.  
 
Consider a neutrino in a superposition of $n$ mass eigenstates  with masses $m_i$, where $i=1,2 \ldots n$.  Its wave function may be written 
\be \psi(x,t)=\begin{pmatrix}{\psi_1(x,t)}\\{\psi_2(x,t)}\\ \vdots\\ \psi_n(x,t)\end{pmatrix}\ .\ee 
Each component of this wavefunction  may be written as a wavepacket, evolving according to 
\be\psi_i(x,t)=\int dp f_i(p) e^{i(p x -E_i(p) t)}\ , \ee where   the momentum wavefunctions $f_i$ are any square integrable functions of $p$, $E_i(p)= \sqrt{p^2+m_i^2}$, and we use units with $\hbar=c=1$. 
The probability density of detecting   flavor $b$ at spacetime point  $(x,t)$ is proportional to
\be\left|\sum_i V_{b i}^* \psi_i(x,t)\right|^2\ , \ee where $V_{bi}^* $ is the amplitude for mass eigenstate $i$ to produce flavor $b$.   In the typical treatment, it is assumed that $V_{b i}\propto U_{bi}$, where $U$ is the  unitary matrix which implements the transformation between the mass and flavor eigenstate bases. Furthermore, it is assumed that initially \be\psi_i(x,0)=U_{ai}f(x)\ , \ee so that initially the  neutrino is in a pure flavor $a$ eigenstate. Note that the approximation that the different mass components initially have exactly the same wavefunction is not exactly true, since the kinematics will imply that different mass components are produced with different momenta distributions. However for the typical case with ultralight neutrinos  this is a good approximation.

Although the exact shape of the wavepacket is not important, for simplicity, we take the initial  
  wavepacket to be a   gaussian which has the particle found in the vicinity of $x=0$ at time $t=0$.   The momentum  for each mass component centered around $p_i$, with spread $1/d$. 
We take  \be f_i(p)= N_i e^{-d^2(p-p_i)^2/2}   \ , \ee where $N_i$ is a normalization constant, $1/d$ is the momentum uncertainty and $d$ is the initial position uncertainty.  For  flavor oscillations,   $d$ should be   small compared with the neutrino oscillation wavelengths  
\be \lambda_{ij}(p)\equiv {4 p\over |m^2_i-m^2_j| \pi  }\ , \ee  and the   momentum uncertainty $1/d$  should be negligible compared with the experimental resolution. Kinematics  of the production will require that $p_i$ is a function of $i$, however any measurement of the momentum with enough precision to distinguish   two mass eigenstates $i$ and $j$ would require a position uncertainty larger than the corresponding  neutrino oscillation wavelength $\lambda_{ij}$. 
We will assume that for for all $i,j$
\be\label{dconstraint}|p_i-p_j|\ll(1/d) \ll p_i \ . \ee  These bounds are typically extremely well satisfied in realistic experimental situations for the known ultralight neutrinos.

At a given time $t$ the location $x_i(t)$ of the center of each mass component of the wavepacket moves with group velocity $v_i$, \be x_i(t)\equiv v_i t={dE_i \over d p} t\bigg|_{p=p_i}={p_i\over\sqrt{p_i^2+m_i^2}} t\approx t\left(1-{m^2_i\over2 p_i^2}\right)\ . \ee
The wave packet will also spread out in space as it moves. We may neglect this spread for a time $t$ satisfying
\be\label{tconstraint}{m_i^2 t\over 2 p_i^2}\ll d\ .\ee   
In  order to measure the momentum with enough precision to distinguish   two mass eigenstates would require a position uncertainty larger than the neutrino oscillation wavelength.  For instance for   neutrino energies of order 1 GeV  and masses of order $0.1$eV/c$^2$,  \Eref{dconstraint} and \Eref{tconstraint} give a very weak constraint on the time,   $t\ll 10^{22}  km/c$.  On the other hand, for oscillations involving 500 MeV neutrinos which mix with a sterile fermion of mass 20 keV, the oscillation  wavelength would be $\sim 0.2  \mu$m  and, assuming the wavepackets are localized more precisely than this, the different mass components of the wavepackets would    separate in space after a maximum time of about 500 m$/c$. Thus even in principle oscillations involving a 20 keV sterile neutrino would not occur for 500 MeV  neutrinos in the MiniBooNE detector.

For ultralight neutrinos, although the different  mass components of the  wavepackets may be  treated as moving with the same group velocity over reasonable timescales,   the relative phase between the wavepackets  will oscillate with a wavelength $\lambda$, leading to flavor  oscillation.   

Specializing to the 2 neutrino case for simplicity of presentation, the probability of finding flavor $a$  at position and time $(x,t)$ is  proportional to
\be P(x,t)=\left|  \int dp \cos\theta N_1 e^{-d^2(p-p_1)^2/2}  e^{i(p x -E_1(p) t)}  +\sin\theta N_2 e^{-d^2(p-p_2)^2/2}  e^{i(p x -E_2(p) t)  } \right|^2 \ .\ee

For 2 flavor oscillations of ultralight neutrinos we may take $N_1=\cos\theta N, N_2=\sin\theta N$. Then
\bea P(x,t)&\approx &N^2 \bigg|  \int dp e^{ip(x-t)}\big(\cos^2\theta  e^{-d^2(p-p_1)^2/2}  e^{ -im_1^2 t\over 2 p} \nn \\ && +\sin^2\theta   e^{-d^2(p-p_2)^2/2}  e^{-im_2^2 t\over 2 p  }\big) \bigg|^2 \\
P(x,t)&\approx &N^2\bigg|  \int dp e^{ip(x-t)}e^{ -im_1^2 t\over 2 p}e^{-d^2(p-p_1)^2/2}   \big(\cos^2\theta  \nn \\ && \quad  +\sin^2\theta   e^{-d^2\left((p_1-p_2)(2 p-p_1-p_2)/2\right)}  e^{-i(m_2^2-m_1^2) t\over 2 p  }\big) \bigg|^2 \ .\eea
We assume $$|p_1-p_2|\ll (1/d)$$ (for fixed energy, this is the same as the assumption $d\ll\lambda$), and note that we get very small  contribution to the integral over $p$ due to destructive interference between different components of the wavefunction unless $|p-p_1|<1/d$. It is therefore reasonable to approximate
\be e^{-d^2\left((p_1-p_2)(2 p-p_1-p_2)/2\right)} \approx 1\ee and get
\be P(x,t)\approx N^2\left|  \int dp e^{ip(x-t)}e^{ -im_1^2 t\over 2 p}e^{-d^2(p-p_1)^2/2}   \left(\cos^2\theta    +\sin^2\theta    e^{-i(m_2^2-m_1^2) t\over 2 p  }\right) \right|^2 \ . \ee

The condition 
${m_i^2 t\over 2 p_i^2}\ll d $ combined with  the fact that  the dominant contribution to the integral has $|p-p_1|<1/d$ allows us to approximate
\be e^{-i(m_2^2-m_1^2) t\over 2 p }\approx e^{-i(m_2^2-m_1^2) t\over 2 p_1 }\ . \ee  This factor then may be taken outside the integral, allowing us to write
\bea P(x,t)&\approx &N^2\bigg|  \int dp e^{ip(x-t)}e^{ -im_1^2 t\over 2 p} e^{-d^2(p-p_1)^2/2}\bigg|^2\bigg|\big(\cos^2\theta    \nn \\ && \quad    +\sin^2\theta     e^{-i(m_2^2-m_1^2) t\over 2 p_1  }\big) \bigg|^2\\
 P(x,t)&\approx &N^2\left|  \int dp e^{ip(x-t)}e^{-d^2(p-p_1)^2/2}\right|^2 \left(1-    \sin^2(2\theta)\sin^2\left({(m_2^2-m_1^2) t\over 4 p_1  }\right)\right)  \ . \eea

Due to the rapidly oscillating $e^{ip(x-t)}$ factor, cancellations between different phase components of this wave function will give a small probability of finding a neutrino of either flavor  except when 
$$|x-t|< d\ ,$$ so oscillations in time will correspond to oscillations in space.  
 Note that have not assumed  that that the different mass eigenstates have either the same momentum  or the same energy expectation value, although we have used  a small neutrino mass approximation.
 
Recently there has been interest in whether the effects of entanglement of the neutrino with other decay products can affect the neutrino oscillation formulae \cite{Cohen:2008qb,HamishRobertson:2010xr,Kayser:2010bj,Akhmedov:2010ua}. To examine this issue for the 2 flavor case, we consider a muon neutrino beam produced from pion decay, which yields
a 2 particle state with wave function
\be\psi(x,y,t)={{\psi_1(x,y,t)}\choose{\psi_2(x,y,t)}}\ . \ee
Here $|\psi_i(x,y,t)|^2$ is the probability   density to  find neutrino mass eigenstate $i$ at $x$  and a muon at $y$. The probability to find neutrino flavor $a$ at $x$ and a muon at $y$  is \be|\cos\theta  \psi_1(x,y,t)+\sin\theta\psi_2(x,y,t)|^2  \ . \ee If $\psi(x,y,t)$  does not factorize into $\chi(y,t)\xi(x,t)$ we say the muon and the neutrino are in an entangled state.  To find the probability of finding neutrino with flavor $a$ at $x$, independent of where the muon is,  we integrate over $y$ to get
\be P(x,t) =\int dy|\cos\theta  \psi_1(x,y,t)+\sin\theta\psi_2(x,y,t)|^2  \ . \ee   Neglecting interactions, each of the wavefunctions $ \psi_i $ satisfies the Schrodinger equation for 2 free relativistic particles, that is a general solution has
\be \psi_i(x,y,t)=\int dq \int dp f_i(p,q) e^{i(px+qy-(E_i(p) + \tilde{E}(q))t} \ee where 
\be \tilde{E}(q)=\sqrt{q^2+M_\mu^2}\ , \ee and $\tilde{E}$ does not require an $i$ index because the muon only comes in 1 mass eigenstate. The fact that the expectation value of the momentum of the muon depends on which neutrino eigenstate is produced is reflected in $f_1\ne f_2$. This is the most general possible solution. To get a wave packet for each particle, with both  localized in the vicinity of the origin  at $t=0$, we choose $ f_i(p,q)$ to be square integrable functions which do not contain any   rapid  change of phase as $p,q$ are varied. A  product of gaussians  or any similarly shaped function centered around $(p_i, q_i)$ will do, with width in both directions greater than $1/\lambda$, consistent with  position uncertainty   less than $\lambda$. As before, to get neutrino oscillations, we assume ${m_i^2 t\over 2 p_i^2}\ll d\ll\lambda$, and $|p_1-p_2| \sim |q_1-q_2|\sim {\cal O} (1/\lambda)\sim {\cal O}(|m_1^2-m_2^2|/p_i)$. 

The probability density of finding a muon flavor neutrino at $x,t$ is
\bea P(x,t) &=&\int \!\!\!\!\int  \!\!\!\!\int  \!\!\!\!\int  \!\!\!\!\int  \!dy dq dq' dp dp'\big(\cos\theta f_1(p,q) e^{-i E_1(p) t}+\sin\theta f_2(p,q)e^{-i E_2(p) t}\big)e^{i(xp+yq-\tilde{E}(q)t)}\nn \\ && \quad\times\big(\cos\theta f_1^*(p',q') e^{i E_1(p') t}+\sin\theta f_2^*(p',q')e^{i E_2(p') t} e^{-i(xp'+yq'-\tilde{E}(q')t)}\big) \ . \eea We can do the $y$ integral, which will give a delta function $\delta(q-q')$. We then can do the $q'$ integral. Note that when we do this, the dependence on the energy of the muon goes away, and we are left with 
\bea  
P(x,t)& = & \!\!\!\int \!\!\!\! \int  \!\!\!\!\int   \!\!\!dq  dp dp'\big(\cos\theta f_1(p,q) e^{-i E_1(p) t}+\sin\theta f_2(p,q)e^{-i E_2(p) t})   (\cos\theta f_1^*(p',q) e^{i E_1(p') t}\\ && \quad+\sin\theta f_2^*(p',q)e^{i E_2(p') t})e^{ix(p-p')} \big) \nn \\   
&=& \!\!\!\int \!\!\!\! \int  \!\!\!\!\int   \!\!\!dq  dp dp'\big(\cos^2\theta(f_1(p,q)f_1(p',q)^* e^{i(x(p-p')-(E_1(p)-E_1(p'))t)}\\ && + \sin^2\theta f_2(p,q)f_2(p',q)^* e^{i(x(p-p')-(E_2(p)-E_2(p'))t)}\\  \nn
&&  +\sin(2\theta) \Re(f_1(p,q)f_2^*(p',q)e^{ix(p-p')-(E_1(p)-E_2(p'))t}) \big) \nn \\ \nn
&\approx& \!\!\!\int \!\!\!\! \int  \!\!\!\!\int   \!\!\!dq  dp dp'\big(\cos^2\theta(f_1(p,q)f_1(p',q)^* e^{i((x-t)(p-p')+(m_1^2 (1/(2p)-1/(2p')))t)} \\
&&+ \sin^2\theta f_2(p,q)f_2(p',q)^* e^{i((x-t)(p-p')+(m_2^2 (1/(2p)-1/(2p')))t)}  \\ \nn
&& +\sin(2\theta) \Re(f_1(p,q)f_2^*(p',q)e^{i(x-t)(p-p')+(m_1^2/(2 p) -m_2^2/(2p'))t}) \big)\ . \nn
\eea Note that as before the location of the expected position in space is correlated with the time, due to an oscillating phase factor  giving destructive interference in the integral unless $x\approx t\pm d/2$. We can therefore consider  flavor oscillations in time, and perform an integral over $x$. Since the center of the wavepacket moves with constant velocity, oscillations in time will correspond to observation of  oscillations in space.

We can find the probability as a function of $t$ to detect  a neutrino of flavor $a$  
\be P(t)=\int^{\infty}_{-\infty} dxP(x,t) \ .\ee The integral over $x$ will provide a delta function $\delta(p-p')$ and so we can also do the integral over $p'$. The result is
\be P(t)\approx \!\!\!\int \!\!\!\! \int  \!\!dq dp \big(\cos^2\theta|f_1(p,q)|^2 + \sin^2\theta| f_2(p,q)|^2  +\sin(2\theta) \Re(f_1(p,q)f_2^*(p,q)e^{i(m_1^2-m_2^2) t/(2 p)}) \big)\ . \ee
Note that the last term, which represents the interference, is now the only term which oscillates in time and it oscillates at the canonical rate. Its size, which gives the amplitude of the oscillations,  may  be reduced by the entanglement, which reduces the overlap of the momentum space wavefunctions for different mass components.    For sufficiently light neutrinos the difference in the central values of the momenta in the momentum space wavefunctions will be much less than the intrinsic uncertainty in the parent particle energy due to its finite lifetime and the effect of the entanglement can be neglected.    

For sufficiently large mass differences the interference terms between different mass components of the neutrino wavefunction will be suppressed due to the near orthogonality of these components, and  the amplitude of oscillations will be suppressed, without affecting the average amount of flavor change. 
\medskip

\noindent {\bf Acknowledgments}
This work was partially supported by the DOE under contract DE-FGO3-96-ER40956. I thank Gerry Garvey, Bill Louis and Jon Walsh for discussions and  Janet Conrad, Georgia Karagiorgi, Belen Gavela and Neal Weiner for correspondence. I thank  the Institute for Nuclear Theory and Wick Haxton, Boris Kayser, Bill Marciano,  and Aldo Serenelli for organizing   the stimulating  Long-Baseline Neutrino Program.
\bibliography{neutrino}
\bibliographystyle{apsrev}
\end{document}